\documentclass[a4paper,10pt]{article}
\usepackage{pos}

\title{Efficient numerical methods for Anisotropic Diffusion of Galactic Cosmic Rays}
\ShortTitle{Numerical Methods for Anisotropic Diffusion}

\author*[a]{Pranab J. Deka}
\author[b]{Ralf Kissmann}
\author[a]{Lukas Einkemmer}

\affiliation[a]{University of Innsbruck, Department of Mathematics,
                Technikerstrasse 13, Innsbruck, Austria, A-6020}

\affiliation[b]{University of Innsbruck, Institute for Astro- and Particle Physics,
                Technikerstrasse 25, Innsbruck, Austria, A-6020}

\emailAdd{pranab.deka@uibk.ac.at}

\abstract{
Anisotropic diffusion is imperative in understanding cosmic ray diffusion across the Galaxy, the heliosphere, and the interplay of cosmic rays with the Galactic magnetic field. This diffusion term contributes to the highly stiff nature of the cosmic ray transport equation. To conduct numerical simulations of time-dependent cosmic ray transport, implicit integrators (namely, Crank--Nicolson (CN)) have been traditionally favoured over the CFL-bound explicit integrators in order to be able to take large step sizes. We propose exponential methods to treat the linear anisotropic diffusion equation in the presence of advection and time-independent as well as time-dependent sources. These methods allow us to take even larger step sizes that can substantially speed-up the simulations whilst generating highly accurate solutions. In our subsequent works, we will use these exponential solvers in the \textsc{Picard} code to study anisotropic cosmic ray diffusion and we will consider additional physical processes such as continuous momentum losses and reacceleration.
}

\FullConference{%
The 38th International Cosmic Ray Conference (ICRC2023)\\
  26 July -- 3 August, 2023\\
  Nagoya, Japan}

\begin{document}
\maketitle

\section{Introduction}

The propagation of highly energetic charged particles in the Galaxy, known as cosmic rays (CRs) can be mathematically described by the following equation \citep{Strong07}:
\begin{multline}
    \frac{\partial u}{\partial t} - \nabla \cdot (\mathcal{D} \nabla u) + \nabla \cdot (\Vec{a}u) - \frac{\partial}{\partial p} \left(p^2 D_{pp} \frac{\partial}{\partial p} \left(\frac{u}{p^2}\right)\right) + \frac{\partial}{\partial p} \left(\dot{p}u- \frac{p}{3} (\nabla \cdot \vec{v})u\right) + \left(\frac{1}{\tau_f} + \frac{1}{\tau_r}\right)u\\ = S(\vec{r}, p, t),
    \label{eq:cr_full}
\end{multline}
where $u \equiv u(\vec{r}, p, t)$ corresponds to the CR density. CRs are transported from their sources to different parts of the Galaxy primarily by means of diffusion. Anisotropic diffusion of CR particles becomes important when the magnitude of the regular (ordered) magnetic field, locally, becomes (significantly) stronger that of the turbulent fields. The significance of anisotropy of CR diffusion and the interplay physics between (anisotropic) CR diffusion and the Galactic magnetic field has been poorly understood owing to the paucity of CR observational data as well as the lack of our knowledge of the magnetic field of our Galaxy.

The \textsc{Picard} code \cite{Kissmann14} was developed to numerically solve the CR transport equation. Although it has been optimised to solve the steady-state transport equation, it is well-suited to handle time-dependent problems. We propose efficient solvers for the anisotropic diffusion problem which will be used in our future work to study and efficiently treat anisotropic CR diffusion in the Galaxy using \textsc{Picard}. 

Since CR transport is a diffusion-dominated problem, we consider a simplified version of Eq. \eqref{eq:cr_full}. The two-dimensional anisotropic diffusion equation in the presence of advection and sources reads
\begin{equation}
    \frac{\partial u}{\partial t} = \nabla \cdot (\mathcal{D} \, \nabla \, u) + \nabla \cdot (\vec{a} u) + S(x, y, t),
    \label{eq:anidiff}
\end{equation}
where $u \equiv u(x, y, t)$, $\vec{a}$ is the advection velocity, and $S(x, y, t)$ corresponds to spatially- and temporally-varying source(s). The diffusion tensor can be written as
\begin{equation*}
    \mathcal{D} =   \begin{bmatrix}
                        D_{xx}(x, y) \quad     D_{xy}(x, y)	\\
                        D_{yx}(x, y) \quad     D_{yy}(x, y) \\
                    \end{bmatrix},
\end{equation*}
where $D_{xx}(x, y)$ and $D_{yy}(x, y)$ are the diffusion coefficients along X-- and Y--directions, respectively, and $D_{xy}(x, y)$ and $D_{yx}(x, y)$ are the off-diagonal terms that give rise to the mixed derivatives in the diffusion equation.


\section{Computing the matrix exponential and exponential-like functions: Leja method}

We propose the use of the exponential quadrature methods \cite{Ostermann10} in combination with the method of polynomial interpolation at Leja points \cite{Caliari04} to integrate a linear differential equation with a time-dependent forcing term, here, Eq. \eqref{eq:anidiff}. The general form of an exponential quadrature rule reads
\begin{equation}
    u^{n+1} = u^n + \Delta t \sum_{i=1}^s b_i(\mathcal{A} \Delta t) (\mathcal{A}u^n + S(t^n + c_i \Delta t)),
    \label{eq:exp_quad}
\end{equation}
where $\mathcal{A} \, (= \nabla \cdot (\mathcal{D} \, \nabla (\cdot)) + \nabla \cdot (\vec{a}))$ is the underlying matrix, $s$ corresponds to the number of stages, $b_i$ are the weights, and $c_i$ are the quadrature nodes. The weights $b_i$ are given as \cite{Ostermann10}
\[b_1(z) = \varphi_1(z) \]
and 
\begin{align*}
    b_1(z) & = \frac{c_2}{c_2 - c_1} \varphi_1(z) - \frac{1}{c_2 - c_1} \varphi_2(z), \\
    b_2(z) & = \frac{-c_1}{c_2 - c_1} \varphi_1(z) + \frac{1}{c_2 - c_1} \varphi_2(z)
\end{align*}
for $s = 1$ and $s = 2$, respectively. Choosing $s = 1$ and $c_1 = 0.5$, one arrives at the second-order (in time) exponential midpoint rule (Eq. \eqref{eq:exp_midpoint}), given as
\begin{equation}
    u^{n+1} = u^n + \varphi_1(\mathcal{A} \Delta t) \left(\mathcal{A}u^n + S \left(t^n + \frac{1}{2} \Delta t \right) \right) \Delta t.
    \label{eq:exp_midpoint}
\end{equation}
The fourth-order (in time) exponential Gauss quadrature can be obtained by choosing $s = 2$, $c_1 = \frac{1}{2} \left(1 - \frac{1}{\sqrt{3}}\right)$, and $c_2 = \frac{1}{2} \left(1 + \frac{1}{\sqrt{3}}\right)$ in Eq. \eqref{eq:exp_quad}, which then reads 
\begin{align}
    u^{n+1} = u^n & + \varphi_1(\mathcal{A} \Delta t) \left(\mathcal{A}u^n + \left(\frac{\sqrt{3} + 1}{2} \right) S(t^n + c_1 \Delta t) + \left(\frac{1 - \sqrt{3}}{2} \right) S(t^n + c_2 \Delta t)\right) \Delta t \nonumber \\ 
    & + \sqrt{3} \varphi_2(\mathcal{A} \Delta t) (S(t^n + c_2 \Delta t) - S(t^n + c_1 \Delta t)) \Delta t.
    \label{eq:exp_quad_Gauss}
\end{align}
The $\varphi_l(z)$ functions are defined recursively as $\varphi_{l+1}(z) = \frac{1}{z} \left(\varphi_l(z) - \frac{1}{l!} \right)$, with $\varphi_0(z) = \exp(z)$ being the matrix exponential. We evaluate the sum of the $\varphi_l(z)$ functions in Eq. \eqref{eq:exp_quad_Gauss} as the exponential of an augmented matrix \cite{Sidje98}. This results in a significant reduction of the computational cost.

We evaluate these $\varphi_l(z)$ functions by interpolating them as polynomials on Leja points. As CR propagation in the Galaxy is a diffusion-dominated problem, we consider only diffusion-dominated cases, i.e. the eigenvalues of $\mathcal{A}$  are large and negative - they lie predominantly on the real axis. We approximate the largest (in magnitude) real eigenvalue, say $\alpha$, using the Gershgorin's disk theorem, and assuming the smallest eigenvalue to be $0$, scale and shift the set of eigenvalues onto the set of Leja points ($\xi$) on the arbitrary interval $[-2, 2]$. The scaling ($c$) and shifting ($\gamma$) factors are computed as $c = \alpha/2$ and
$\gamma = -\alpha/4$, respectively. Now, we interpolate the function $\exp((c + \xi \gamma) \Delta t)$ or $\varphi((c + \xi \gamma) \Delta t)$ on the set of Leja points, and the $(m + 1)^\mathrm{th}$
term of the polynomial, so formed, is given by
\begin{align*}
    p_{m+1} & = p_m + d_{m+1} \, y_{m+1}, \\
    y_{m+1} & = y_m \times \left(\frac{z - c}{\gamma} - \xi_{m} \right),
\end{align*}
where $d_m$ are the coefficients of the polynomial determined using divided differences. We use the open-source \texttt{LeXInt} library \cite{Deka23a} to evaluate the matrix exponential and $\varphi_l(z)$ functions using the Leja method.


\section{Test examples of Anisotropic Diffusion}

To study the performance of our proposed  method, we consider three different magnetic field configurations. The first configuration is the concentric circles or rings \cite{Sharma07, Crouseilles15, Pakmor16b, Hopkins17}, where the diffusion coefficients depend on the spatial coordinates as
\begin{equation}
    \mathcal{D}_R = \begin{bmatrix}
                        y^2  &    -xy   \\
                        -xy  &     x^2  \\
                    \end{bmatrix}.
    \label{eq:B_ring}
\end{equation}
As the Milky Way is a spiral Galaxy, we consider two different spiral magnetic field configurations given as follows:
\begin{equation}
    \mathcal{D}_{S1} =  \begin{bmatrix}
                            (x + y)^2  &    -x^2 - xy  \\
                            -x^2 - xy  &          x^2  \\
                        \end{bmatrix}
\qquad \qquad
    \mathcal{D}_{S2} = \begin{bmatrix}
                            (x + 4y)^2  &    -x^2 - 4xy \\
                            -x^2 - 4xy  &          x^2  \\
                        \end{bmatrix}.
    \label{eq:B_spiral}
\end{equation}

We consider several scenarios to show the superiority of our proposed solvers over the commonly-used solver for the time-dependent CR transport equation - the CN solver.
\begin{itemize}

    \item \textbf{Case I}: Anisotropic diffusion along a ring, where the magnetic field configuration is given by Eq. \eqref{eq:B_ring}.

    \item \textbf{Case II}: Anisotropic diffusion along a spiral, where the magnetic field configuration is given by Eq. \eqref{eq:B_spiral} (spiral 2), with advection ($\vec{a} = \hat{i} + \hat{j}$) and time-independent sources, given as,
    \begin{equation}
        S(x, y) = 0.1 + 30 \exp\left(-\frac{(x + 0.6)^2 + (y - 0.75)^2}{2.5 \cdot 10^{-2}}\right) + 40 \exp\left(-\frac{(x - 0.75)^2 + (y + 0.8)^2}{3 \cdot 10^{-2}} \right).
        \label{eq:TIS}
    \end{equation}

    \item \textbf{Cases III and IV}: Anisotropic diffusion along two different spiral magnetic field configurations (Eq. \eqref{eq:B_spiral}), with advection and the following time-dependent sources:
    \begin{align}
        S_A(x, y, t) & = 100 \exp\left(-\frac{(x - 0.25)^2 + (y - 0.25)^2}{1.5 \cdot 10^{-2}}\right) \exp\left(-10 \, |t - 0.1| \right) \nonumber \\
        S_B(x, y, t) & = 70 \exp\left(-\frac{(x - 0.4)^2 + (y + 0.4)^2}{2.5 \cdot 10^{-2}}\right) \exp\left(-5 \, |t - 0.3|\right) \nonumber \\
        S_C(x, y, t) & = 50 \exp\left(-\frac{(x + 0.6)^2 + (y - 0.6)^2}{1.5 \cdot 10^{-2}}\right) \exp\left(-7.5 |t - 0.6| \right) \nonumber \\
        S(x, y, t) & = S_A(x, y, t) + S_B(x, y, t) + S_C(x, y, t).
        \label{eq:TDS}
    \end{align}
    
\end{itemize}

\begin{figure*}
    \centering
    \includegraphics[width = \textwidth]{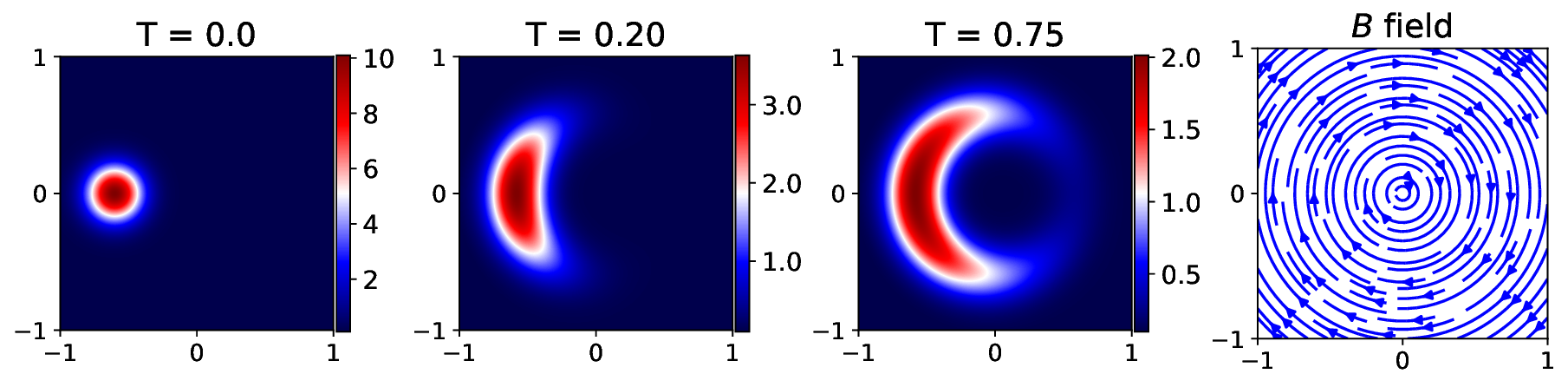} 
    \includegraphics[width = \textwidth]{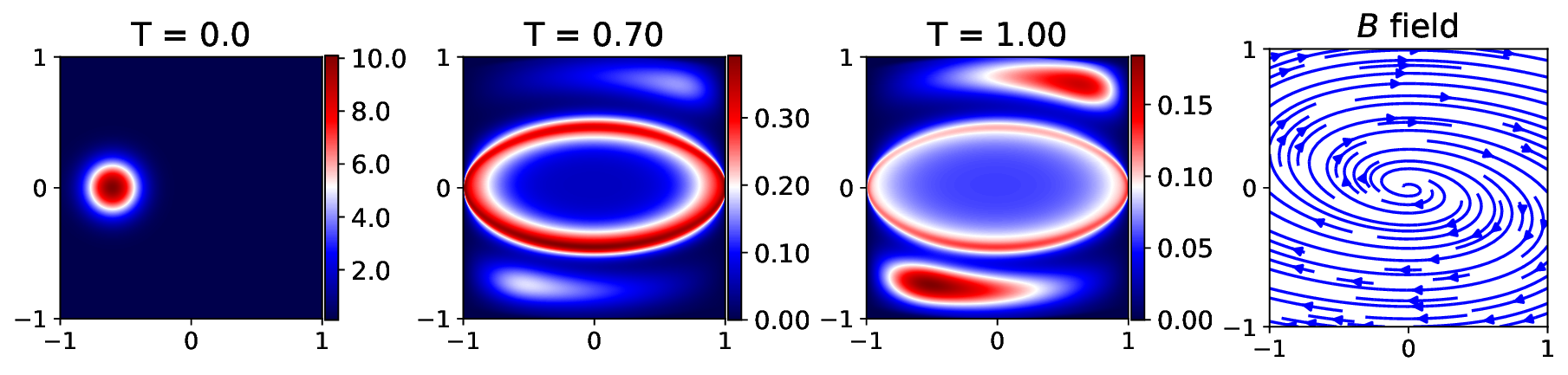} 
    \caption{Anisotropic diffusion of an initial Gaussian distribution along the field lines where the magnetic field configuration is in the form of concentric circles (top) and spiral 2 (bottom). We consider two time-independent sources in the case of diffusion along the spirals.}
    \label{fig:ring}
\end{figure*}

\begin{figure*}
    \centering
    \includegraphics[width = \textwidth]{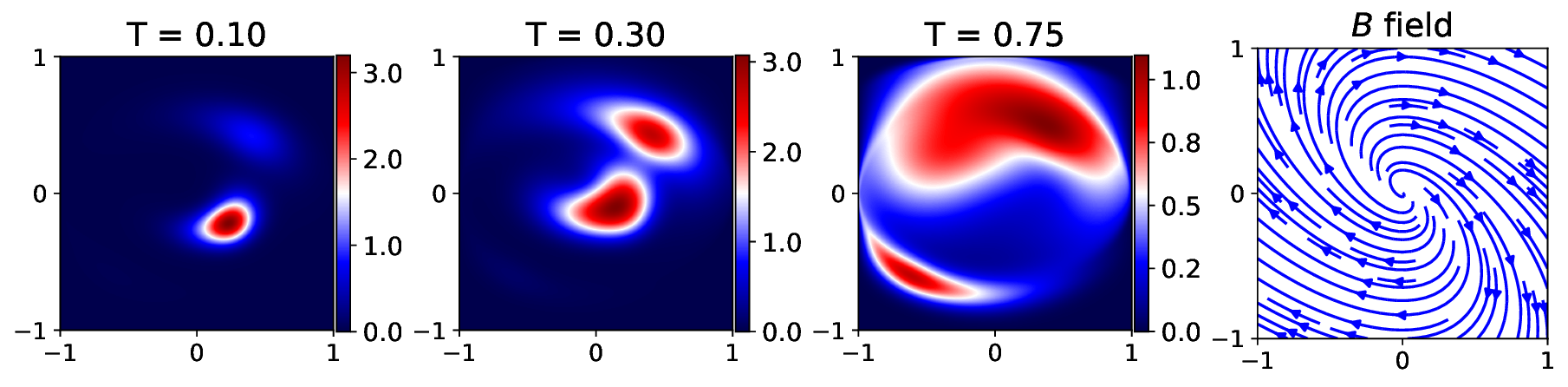} 
    \includegraphics[width = \textwidth]{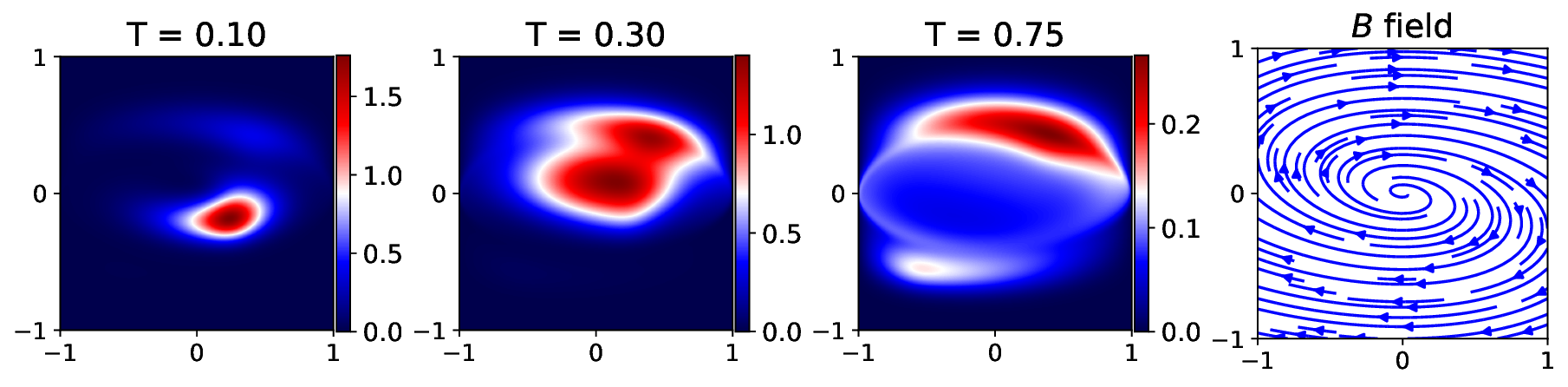}
    \caption{Anisotropic diffusion along spiral magnetic field configuration (spiral 1 - top and spiral 2 - bottom). We consider three time-dependent sources (Eq. \eqref{eq:TDS})  in each of these cases.}
    \label{fig:spiral_TDS}
\end{figure*}

In Fig. \ref{fig:ring}, we illustrate anisotropic diffusion of an initial Gaussian distribution along a ring and along a spiral in the presence of two time-independent sources. One can clearly see the two sources pop up once the initial distribution has been smeared out enough such that it is roughly as intense as the two sources. In Fig. \ref{fig:spiral_TDS}, we consider anisotropic diffusion and advection ($\vec{a} = \hat{i} + \hat{j}$ for spiral 1 and $\vec{a} = \hat{i} + 2\hat{j}$ for spiral 2) of the distribution function with time-dependent sources for two different spiral field configurations. This is reminiscent of CR sources (say, supernovae) injecting CRs at different times and positions in the Galaxy.

\begin{figure*}
    \centering
    \includegraphics[width = \textwidth]{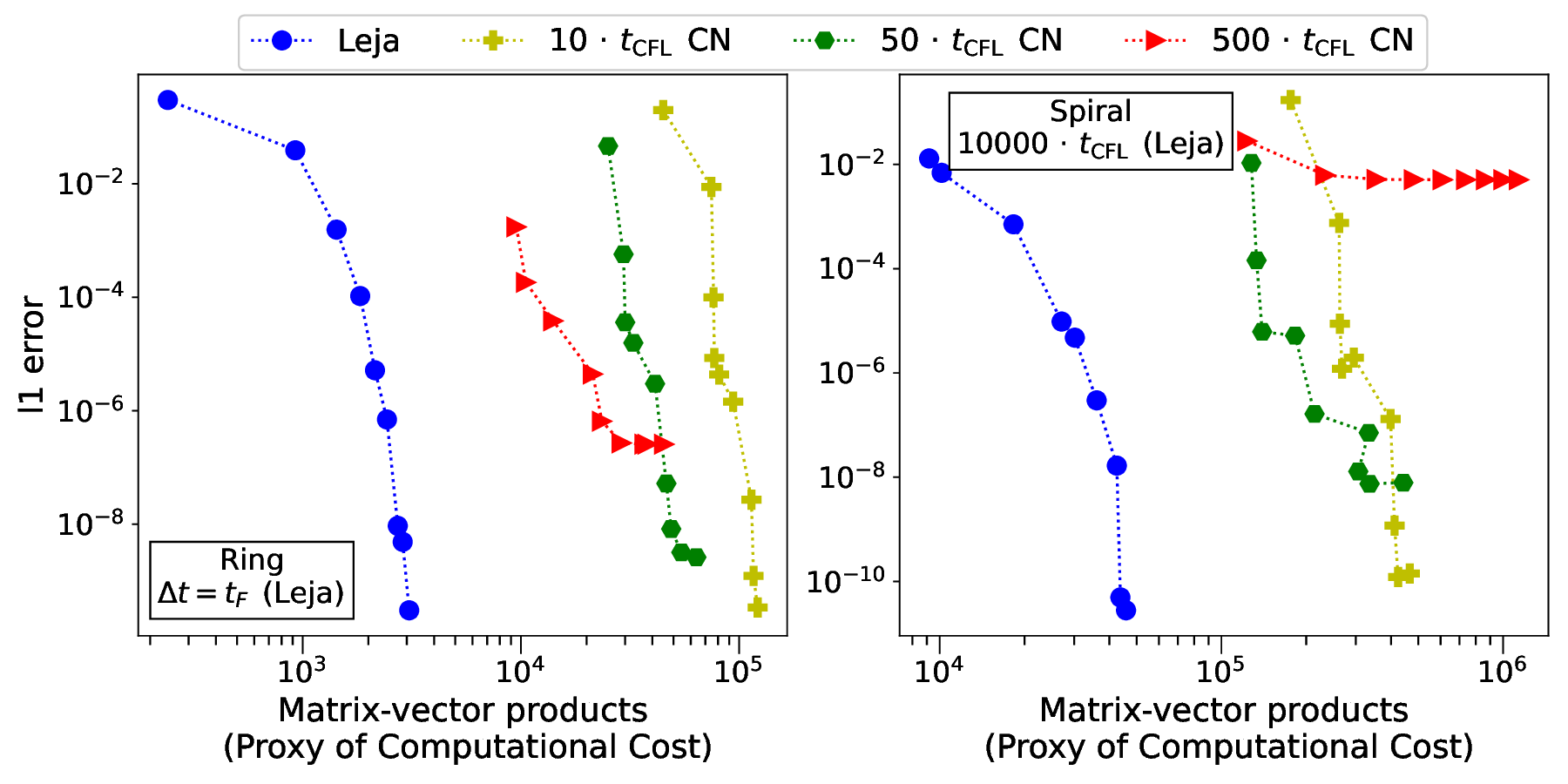}
    \caption{Left: Work-precision diagram, for diffusion along a ring, at T = 0.75 \textcolor{blue}{\cite{Deka23b}}. Right: Work-precision diagram, for diffusion (+ advection + sources) along a spiral, at T = 0.85.}
    \label{fig:TIS_WP}
\end{figure*}

\begin{figure*}
    \centering
    \includegraphics[width = \columnwidth]{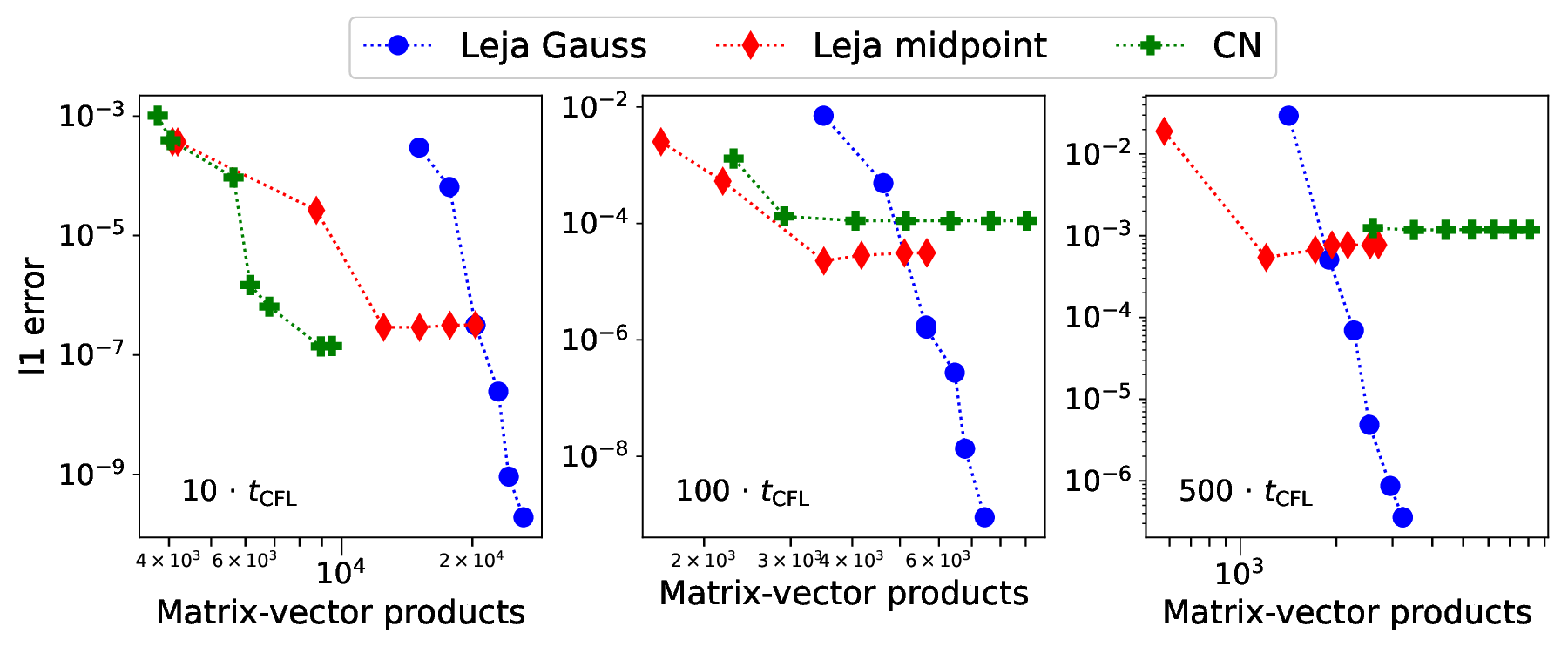} \\
    \includegraphics[width = \columnwidth]{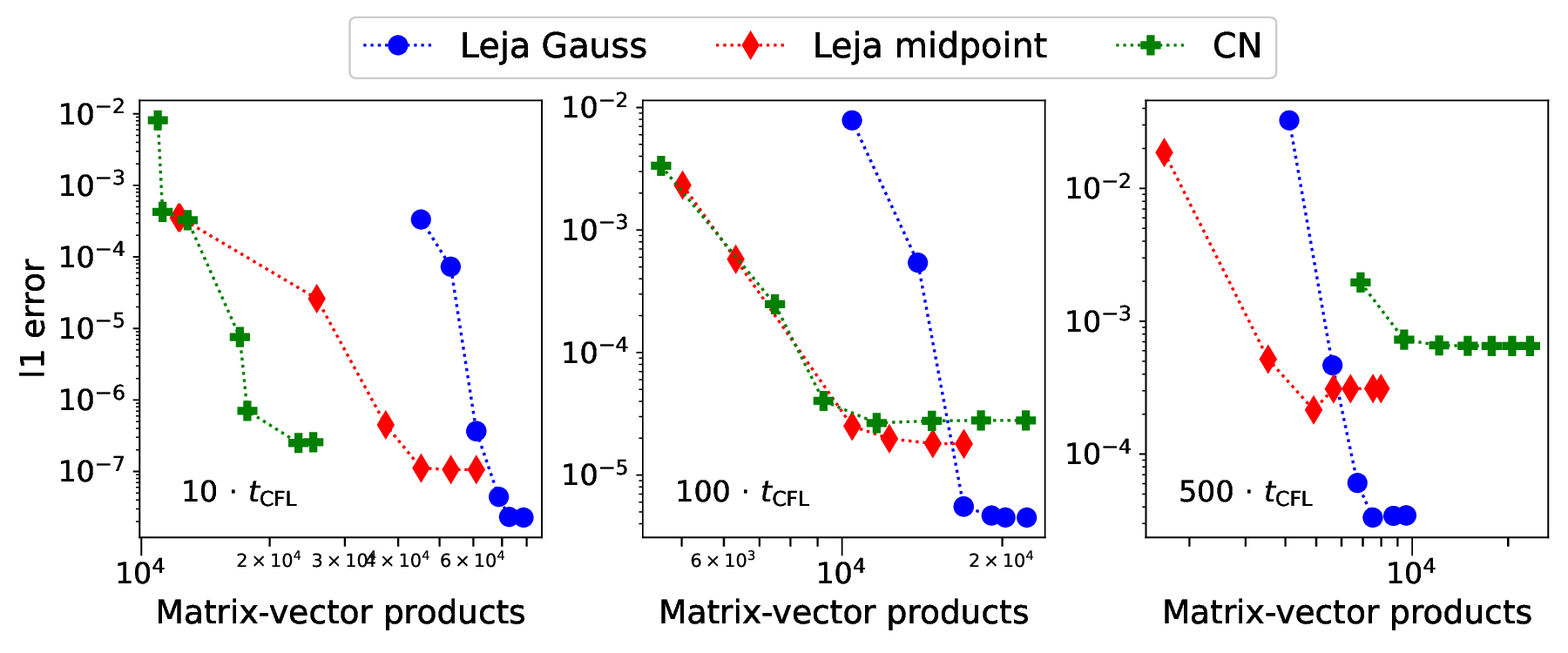}
    \caption{Work-precision diagram, for diffusion along spiral 1, for anisotropic diffusion, advection and time-dependent sources at $T = 0.10$ (top) and $T = 0.30$ (bottom).}
    \label{fig:TDS_WP_S1}
\end{figure*}

\begin{figure*}
    \centering
    \includegraphics[width = \columnwidth]{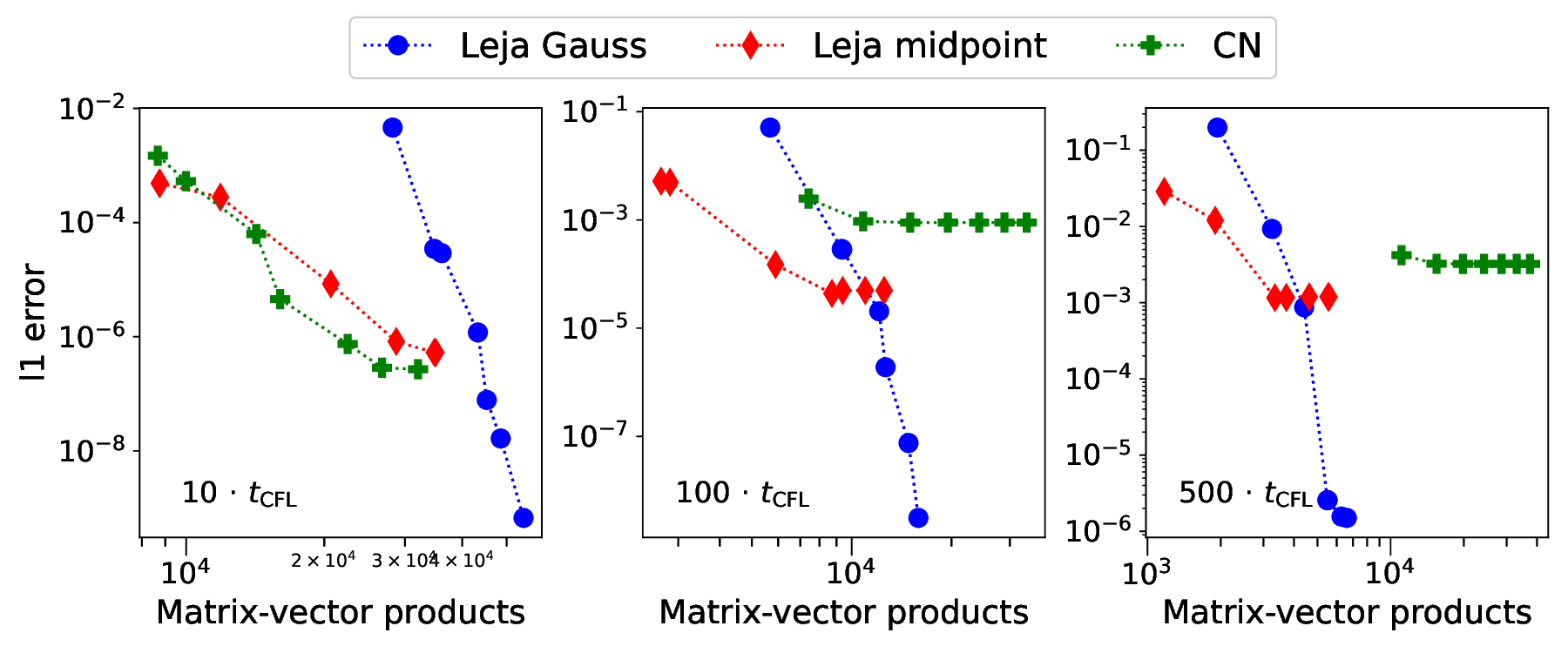} \\
    \includegraphics[width = \columnwidth]{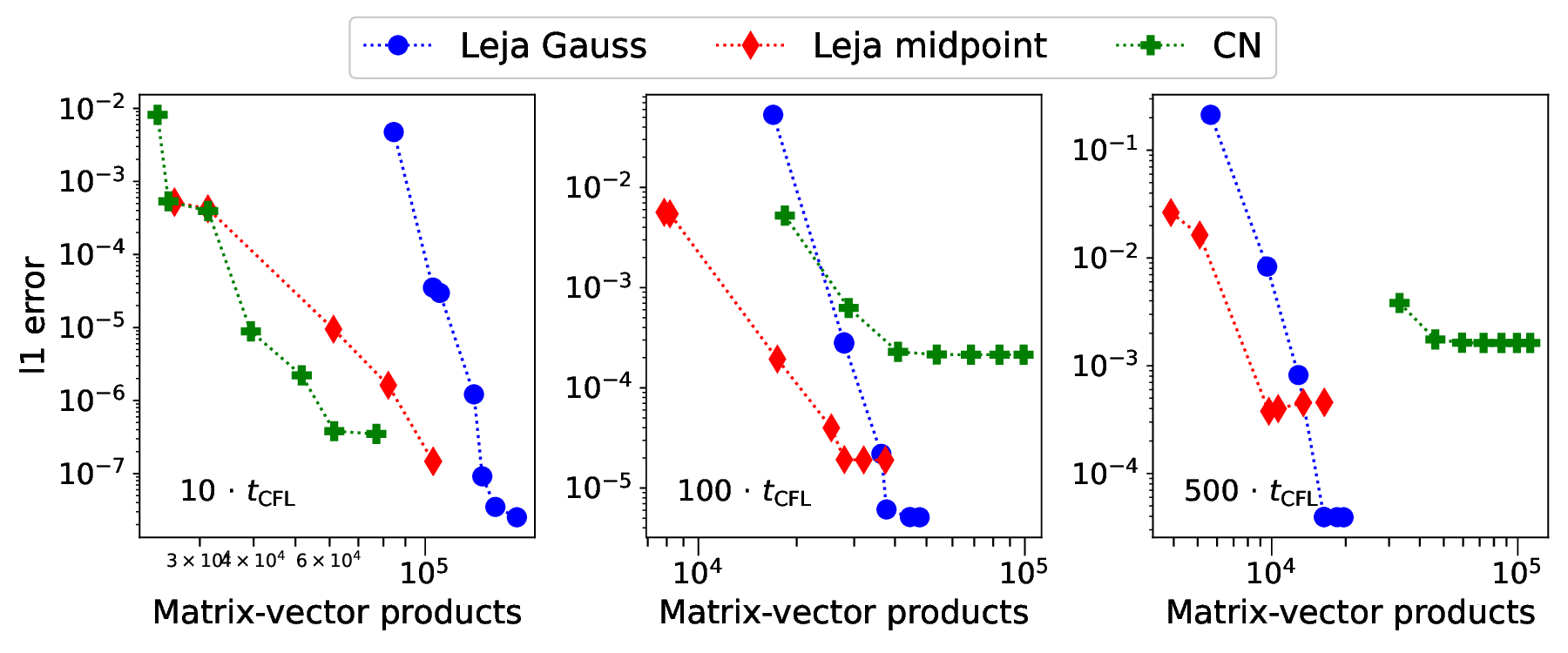}
    \caption{Work-precision diagram, for diffusion along spiral 2, for anisotropic diffusion, advection and time-dependent sources at $T = 0.10$ (top) and $T = 0.30$ (bottom).}
    \label{fig:TDS_WP_S2}
\end{figure*}

Fig. \ref{fig:TIS_WP} contrasts the performance of computing the matrix exponential directly using the Leja method with that of CN for anisotropic diffusion on a ring (at $T = 0.75$) and a spiral (at $T = 0.85$, including advection and time-independent sources). We consider tolerances (i.e., desired accuracy of the solutions) in steps of an order of magnitude from $10^{-2}$ to $10^{-10}$, and plot the resulting error incurred as a function of a proxy of the computational runtime, here, the number of matrix-vector products, as that constitutes the most expensive part of the algorithm. The Leja method (blue circles) is significantly cheaper than the CN solver owing to its ability to take substantially larger step sizes than its counterpart. In the case of diffusion along a ring, one can take step sizes as large as the total simulation time. This results in performance improvements from a factor of 5 for intermediate tolerances up to almost two orders of magnitude for stringent tolerances (highly accurate solutions). Similar results are obtained for diffusion along a spiral. The additional physical phenomena, i.e. advection and sources, imposes practical (or numerical) restrictions on the permissible step sizes. Nevertheless, one is able to choose step sizes as large as $10000$ times the CFL time step size, thereby obtaining speedups of about an order of magnitude.

In the case of time-dependent sources, one can no longer obtain the \textit{exact} solution in time. The time-step sizes are restricted by the relevant physical processes at play. In Figs. \ref{fig:TDS_WP_S1} and \ref{fig:TDS_WP_S2}, we compare the performance of the exponential midpoint (Eq. \eqref{eq:exp_midpoint}) and exponential Gauss (Eq. \eqref{eq:exp_quad_Gauss}) quadrature methods, where the $\varphi_l(z)$ functions are approximated by the Leja method, with the CN method for $T = 0.10$ (top panel) and $T = 0.30$ (bottom panel). Exponential Gauss quadrature (blue circles), being fourth order convergent, is more accurate than exponential midpoint quadrature (red diamonds) and CN (green pluses), for all considered configurations. For small step sizes ($\Delta t = 10 \cdot \Delta t_\mathrm{CFL}$), the CN solver needs fewer iterations per time step, and therefore, proves to be cheaper (by a factor of 2 - 3) than the exponential midpoint method and up to a factor of 4 - 5 than the exponential Gauss quadrature. 

For larger step sizes ($\Delta t = 100 \cdot \Delta t_\mathrm{CFL}$ and $500 \cdot \Delta t_\mathrm{CFL}$), CN suffers from a significant increase in the computational runtime as well as a significant loss in accuracy. Exponential midpoint quadrature suffers a similar fate in accuracy, however, it converges relatively rapidly as compared to the CN solver. This difference in the rate of convergence can range somewhere between a factor of $2$ to $10$ depending on the specifications of the problem under consideration. Both of these methods, being second-order convergent, rapidly reaches a saturation in the error incurred for large step sizes, irrespective of the prescribed tolerance. Exponential Gauss quadrature is clearly superior to both the second-order solvers for large step sizes - it can yield highly accurate solutions, an improvement of a factor of 3 to several orders of magnitude, depending on the simulation time and other parameters, for similar computational effort. 

\section{Conclusions and Outlook}

We have shown that exponential methods have better performance in terms of computational runtimes as well as accuracy than the Crank-Nicolson solver, which has been traditionally used in \textsc{Picard} and \texttt{GalProp} for treating the time-dependent CR transport equation. Even though we have only considered two-dimensional problems in this work, we fully expect that our proposed solvers would show similar improvements in performance, both in terms of computational cost and the accuracy of the solutions, for the full four-dimensional CR transport equation.

In the near future, we will append \texttt{LeXInt} to \textsc{Picard} to study CR transport in the Galaxy using the exponential quadrature methods in combination with the Leja scheme. Yet another approach would be to penalise the anisotropic diffusion term \cite{Deka23b} - the stiff terms would be treated implicitly whilst the non-stiff term would be solved explicitly. We will also consider continuous energy losses as well as reacceleration whilst investigating the impact of anisotropic CR diffusion in the Galactic magnetic field using \textsc{Picard}.


%
%
%

\end{document}